\documentclass[twocolumn,showpacs,preprintnumbers,amsmath,amssymb,superscriptaddress]{revtex4}

\usepackage{graphicx}
\usepackage{dcolumn}
\usepackage{bm}

\begin{document}

\title{Diamagnetic Phase Transition and Phase Diagrams\\in Beryllium}

\author{Nathan Logoboy}

\email{logoboy@phys.huji.ac.il}

\affiliation{Grenoble High Magnetic Field Laboratory, MPI-FKF and
CNRS P.O. 166X, F-38042 Grenoble Cedex 9, France}

\affiliation {The Racah Institute of Physics, The Hebrew University
of Jerusalem, 91904 Jerusalem, Israel}

\author{Walter Joss}
\affiliation{Grenoble High Magnetic Field Laboratory, MPI-FKF and
CNRS P.O. 166X, F-38042 Grenoble Cedex 9, France}

\affiliation {Universit$\acute{e}$ Joseph Fourier, B.P. 53, F-38041e
Grenoble Cedex 9, France}

\date{\today}

\begin{abstract}
The model of diamagnetic phase transition in beryllium which takes
into account the quasi $2$-dimensional shape of the Fermi surface of
beryllium is proposed. It explains correctly the recent experimental
data on observation of non-homogeneous phase in beryllium at the
conditions of strong dHvA effect when the strong correlation of
electron gas results in instability of homogeneous phase and
formation of Condon domain structure.
\end{abstract}

\pacs{75.20.En, 75.60.Ch, 71.10.Ca, 71.70.Di, 71.25.-s; 71.25. Hc; 75.40-s; 75.40.Cx.}
\maketitle

\section{\label{sec:level1}Introduction}

It is well-known \cite{Shoenberg} that the instability in electron
gas due to magnetic interaction between conduction electrons in
diamagnetic metals at high magnetic field and low temperature leads
to formation of non-uniform diamagnetic phase, so-called Condon
domains, which is usually realized as a stripe-domain structure for
plate-like samples. The diamagnetic domains were observed in
beryllium by magnetization measurements \cite{Condon} and in silver
by nuclear magnetic resonance (NMR) \cite{Condon_Walstedt} and by
Hall probes spectroscopy \cite{Kramer1}. They have been observed in
beryllium, white tin, aluminium and lead by muon spin-rotation
($\mu$SR) spectroscopy \cite{Solt1}. The above-mentioned instability
of an electron gas is called diamagnetic phase transition and is
extensively studied, both theoretically and experimentally
\cite{Kramer2}-\cite{Gordon1}.

Recent progress in experiments on observation of Condon domain
structure in silver \cite{Kramer1} and beryllium
\cite{Kramer2}-\cite{Tsindlekht} provides a natural stimulus towards
a more detailed understanding of the properties of strongly
correlated electron gas at the conditions of dHvA effect.
Undoubtedly, silver and beryllium are the most popular metals in
experimental investigation of electron instability leading to
formation of Condon domain structures. The direct observation of
Condon domains in silver by Hall probe spectroscopy \cite{Kramer1}
shows that the temperature and magnetic field dependencies of
non-uniform phase are in good agreement with the theoretically
predicted phase diagrams (see, e.g. \cite{Gordon1}), estimated on
the basis of the Lifschitz-Kosevich-Shoenberg formalism in the first
harmonic approximation \cite{Shoenberg}. However, in case of
beryllium, the standard approach, which gives the satisfactory
results for free or almost free electron gas, fails to describe the
diamagnetic phase transition, and, leads, in particular, to the {\it
underestimated} values for critical temperature $T_{c}$ and magnetic
field $H_{c}$, which is inconsistent with the experimental data
\cite{Solt1}-\cite{Tsindlekht}. Thus, for proper calculation of
amplitude of the dHvA oscillations and constructing the phase
diagrams, the correct topology of Fermi surface has to be taken into
account. Recent experimental observation of irreversible effects in
beryllium by Hall probes in dc field and standard ac method with
various modulation levels, low frequencies and magnetic field ramp
rates \cite{Kramer2}, as well as application of non-linear
techniques \cite{Tsindlekht}, which reveals giant parametric
amplification of non-linear response in Condon domain phase, offer
the possibility to construct the diamagnetic phase diagrams for
beryllium and compare it with theoretical predictions.

The Fermi surface of beryllium was investigated very carefully in
the past \cite{Shoenberg}. Paradoxically, beryllium has the simplest
Fermi surface, because it differs essentially from the free electron
model. The Fermi surface of beryllium consists only of the first and
second zone monster ('coronet') and the third zone 'cigar'. It is
well-established that the dHvA oscillations originates from the
three maximum cross-sections of 'cigar' ('waist' and 'hips') which
are characterized by a very small curvature. The small curvature of
the cylinder like Fermi surface explains relatively high amplitude
of dHvA oscillations in beryllium at magnetic field H applied
parallel to the hexagonal axis, e.g. $H \parallel [0001]$, in
comparing to silver \cite{Kramer1}, where the small amplitude of
dHvA oscillations is explained in the framework of spherical Fermi
surface \cite{Shoenberg} providing the reasonable agreement between
estimated phase diagrams \cite{Gordon1} and experimental data
\cite{Kramer1}. A new experimental results
\cite{Kramer2}-\cite{Tsindlekht} on observation of Condon domain
phase in beryllium stimulate the further development of the theory.
Undoubtedly, for correct explanation of experimental data
\cite{Kramer2}-\cite{Tsindlekht} the anomalously low curvature of
the 'cigar' like part of Fermi surface of beryllium near the extreme
cross-sections ('waist' and 'hips') has to be taken into account.
The modeling of cigar-like part of the Fermi surface of beryllium
\cite{Egorov} by a cylinder, similar to 2D electron gas, describes
correctly the diamagnetic phase diagrams at low range of quantizing
magnetic field ($\le 3 T$ ), but results in the essentially {\it
 overestimated} values of critical parameters (temperature $T_{c}$
and magnetic field $H_{c}$) at higher values of applied magnetic
field. In particular, the models based on approximation of relevant
Fermi surface sheets by cylinder predict the existence of
inhomogeneous phase at the values of external magnetic field $H \sim
10-12 T$ contradicting to the experimental data
\cite{Kramer2}-\cite{Tsindlekht} that show the disappearance of
Condon domain structure above $\sim 6 T$ at the values of Dingle
temperature $T_{D}\approx 2.2 K$. Moreover, the approximation of
Fermi surface of beryllium by cylinder \cite{Egorov} fails to
explain the observed beatings in the dHvA oscillations
\cite{Condon_Walstedt},\cite{Solt1}-\cite{Tsindlekht}, which are the
result of the interference of the signals from two different extreme
cross-sections of Fermi surface \cite{Solt1} with close fundamental
frequencies. Solt \cite{Solt1} shows the serious disagreement of
$\mu$SR data in beryllium with the phase diagrams, calculated in the
framework of standard Lifshitz-Kosevich formula, and necessity to
take into consideration the actual 3D Fermi surface geometry for
beryllium, first, the existence of two different cross-sections
('waist' and 'hips'), and second, the low curvature of Fermi surface
near extreme cross-sections. The model developed in
\cite{Solt1}-\cite{Solt3} is more realistic; it describes correctly
the beating effect in dHvA oscillations in beryllium, and it is
consistent with the previous experiments on observation of Condon
instability \cite{Solt1} at low range of quantizing magnetic field
(till $\sim 3 T$). Unfortunately, the high values of critical
magnetic field $\sim 10 T$ for non-homogeneous phase, estimated in
the framework of model \cite{Solt3}, contradict to recent data on
independent observation of Condon instability in beryllium
\cite{Kramer2}-\cite{Tsindlekht}, which demonstrate the existence of
non-uniform phase in essentially narrow interval of quantizing
magnetic field ($\approx 0-6 T$) depending on estimated Dingle
temperature $T_{D}\approx 2.2 K$. In present paper we develop the
model of slightly corrugated cigar-like Fermi surface for beryllium
and calculate the diamagnetic phase diagrams T-H at different Dingle
temperatures $T_{D}$. The estimated phase diagrams are in good
agreement with the available experimental data on observation of
diamagnetic instability in beryllium \cite{Solt1}-\cite{Tsindlekht}.

The paper is organized as follows. Section I is an introduction. In
Section II we consider the model of slightly corrugated cylinder
Fermi surface of beryllium. The model is based on reliable knowledge
of the dimensions of the relative Fermi surface sheets
\cite{Shoenberg}, as well as its qualitative nature. In Section III
we calculate the diamagnetic phase diagrams for beryllium and
compare the theoretical results with recent observation of electron
instability in beryllium \cite{Kramer2}-\cite{Tsindlekht}. Section
IV is conclusions. Section V contains acknowledgements.

\section{Model}
The oscillatory part of the free energy density which is responsible
for dHvA effect can be written as follows
\cite{Shoenberg}-\cite{Kubler}

\begin{equation} \label{eq:free energy density}
\Omega =\frac{(e\mu_{0}H)^{2}} {(2\pi^{2}c)^{2}m_{c}} R(T,\mu_{0}H,T_{D})\int dk_{z}\cos[2\pi X(k_{z})], \\
\end{equation}
where
\begin{equation} \label{eq:R}
R(T,\mu_{0}H,T_{D})=\frac{\lambda(\mu_{0}H)T\exp {[-\lambda(\mu_{0}H)T_{D}]}}{\sinh {[\lambda(\mu_{0}H)T]}}. \\
\end{equation}
Here, $e$ is the absolute value of electron charge, $c$ is the light
velocity, $k_{B}$ is the Boltzmann constant, $m_{c}$  is the
cyclotron mass,
$\lambda(\mu_{0}H)=2\pi^{2}k_{B}m_{c}c/e\hbar\mu_{0}H$ and
$T_{D}=\hbar/2\pi k_{B}\tau$ is Dingle temperature which is
inversely proportional to the scattering lifetime $\tau$ of the
conduction electron. The cross-sectional area of the Fermi surface
$\mathcal{A}(\mu, k_{z})$ at $k_{z}$ in $\mathbf k$-space is related
to the quantity $X(k_{z})$ by the relationship
\begin{equation} \label{eq:X}
X(k_{z})=\frac{c\hbar \mathcal{A}(\mu k_{z})}{2\pi e \mu_{0}H}-\gamma, \\
\end{equation}
where $\mu$ is the chemical potential and $\gamma$ is the phase
correction (typically, it equals to 1/2).

The integral in Eq.~(\ref{eq:free energy density}) is a
Fresnel-type. Its major contributions come from regions where the
phase is stationary \cite{Shoenberg}, e.g. the cross-sectional area
$\mathcal{A}(\mu, k_{z})$ has maximums or minimums. The usual
procedure of calculation of such an integral consists in expanding
of electron orbit area $\mathcal{A}(\mu, k_{z})$ about the extreme
points and summation of the contributions from different extreme
cross-sections of Fermi surface \cite{Shoenberg}. In case of
beryllium the standard procedure requires considerable modification
due to vanishing low curvature of extreme cross-sections of 'cigar',
when the higher order terms in expansion need to be considered. The
additional disadvantage of standard expansion of $\mathcal{A}(\mu,
k_{z})$ about the extreme points arises from the fact that neither
curvature factor nor its derivatives are available from the
experiment. The last circumstance implies the necessity of
development of model representations of the Fermi surface of
beryllium.

Our consideration is based on the following model representation of
electron orbit area for the 'cigar' like sheet of Fermi surface of
beryllium relevant for observed dHvA oscillations:

\begin{equation}\label{eq:model}
\mathcal{A}(k) = \left\{ \begin{array}{ll}
 1+\delta, & 0\le \mid k \mid \le 1-\beta\\
 1+\delta\sin{\alpha(k-0.5)}, & \textrm{$\beta < \mid k \mid \le 1-\beta$}\\
 1+\delta, & \textrm{$1-\beta < \mid k\mid \le 1+\beta$}\\
 1-\delta \sin{\alpha(k-0.5)}, & \textrm{$1+\beta < \mid k\mid \le
 2+\beta$}\\
  \end{array} \right.
\end{equation}
where $k=k_{z}/\Delta$ is reduced wave vector, $a=\pi/(1-2\beta)$,
$\beta=k_{c}/\Delta$ and $\Delta=0.217~1/A^{^{\circ}}$
\cite{Shoenberg} is the distance in reciprocal space between two
extreme cross sections of 'cigar'. The quantity
$\delta=(\mathcal{A}_{h}-\mathcal{A}_{w})/2\mathcal{A}_{0}$
characterizes the discrepancy between maximal ($\mathcal{A}_{h}$)
and minimal ($\mathcal{A}_{w}$) cross sections of Fermi surface, and
$\mathcal{A}_{0}=(\mathcal{A}_{h}-\mathcal{A}_{w})/2$ is average
cross section area. According to Eq.~(\ref{eq:model}) the extreme
cross sections of 'cigar' are characterized by the same curvature
which is consistent with the data on Fermi surface of beryllium
\cite{Condon}. In calculation of phase diagrams one must take into
consideration the transition region between 'waist' and 'hips'. We
use the approximation of the transition regions by trigonometric
functions (see, Eq.~(\ref{eq:model})), defined in corresponding
intervals through the adjustable parameter $0< \beta <1/2$.
Although, the Fermi surface of beryllium was investigated very
carefully in the past \cite{Shoenberg}-\cite{Condon}, still there is
lack of data concerning the transition between extreme cross
sections. We show that with suitable choice of this parameter the
satisfactory agreement between calculated phase diagrams and
experimental data can be achieved. Before calculation of the phase
diagrams we emphasize that our choice of the representation of the
Fermi surface sheet of beryllium in the form of Eq.~(\ref{eq:model})
is governed by the experimental facts
\cite{Shoenberg}-\cite{Condon_Walstedt} which indicate on the
negligible and equal (or almost equal in the accuracy of
experiments) curvatures of extreme cross sections. The last
circumstance, e.g. negligibility and equality of the 'waist' and
'hips' curvatures, simplifies our task and allows us to neglect the
curvatures in the vicinity $\sim 2\beta$ of the extreme cross
sections with proper choice of the functions describing the
transition between two extreme cross sections.

\section{Results}

The oscillating part of the diamagnetic susceptibility
$\widetilde{\chi}=-\partial
\widetilde{M}=-\partial^{2}\Omega/(\partial B)^{2}$ is obtained by
differentiating Eq.~(\ref{eq:free energy density}) twice with
respect to the magnetic induction $B=\mu_{0}H+M$. Calculating the
integral in Eq.~(\ref{eq:free energy density}) with taking into
account the representation (\ref{eq:model}) and keeping the leading
terms only (see, e.g. \cite{Shoenberg}), one can arrive to the
following expression for the susceptibility
\begin{equation} \label{eq:susceptibility}
\widetilde{\chi}=a \cos{(l^{2}\mathcal{A}_{0}+\psi)}, \\
\end{equation}
where
\begin{equation} \label{eq:reduced amplitude}
a=\frac{4\Delta (\mathcal{A}_{0}\hbar)^{2}}{\pi^{3}m_{c}(\mu_{0}H)^{2}}\mid Q(\mu_{0}H)\mid R(T,\mu_{0}H,T_{D}), \\
\end{equation}
is reduced amplitude of dHvA oscillations, e.g. the ratio between
the amplitude of dHvA oscillations and their period.

The complex function
\begin{eqnarray} \label{eq:Q}
Q=Q_{1}+Q_{2}=\mid Q\mid \exp{j\psi}, \qquad \qquad \nonumber \\
Q_{1}=3\beta \cos{x}+2(1-2\beta) J_{0}(x), \quad Q_{2}=\beta \sin{x}
\end{eqnarray}
where $x=l^{2}\mathcal{A}_{0}\delta$, depends on applied magnetic
field through parameter $l^{2}=c\hbar/e\mu_{0}H$. Here, $J_{0}(x)$
is Bessel function of the first order.

When the amplitude of dHvA oscillations become high enough ($a>1$
Eq.~(\ref{eq:reduced amplitude})) the magnetic interaction between
electrons results in thermodynamic instability of uniform phase and
diamagnetic phase transition into non-uniform phase with formation
of Condon domain structure occurs. The condition
\begin{equation} \label{eq:critical curves}
a(T,\mu_{0}H,T_{D})=1 \\
\end{equation}
defines the critical curves on the plane $T-H$ at different Dingle
temperatures $T_{D}$. The Dingle temperature $T_{D}$ characterizing
the quality of the material can play a crucial role in observation
of electron instability. Thus, for comparing the theoretical results
with experiment it is important to know the values of $T_{D}$ , for
which the Condon domain structure can be observed in principal at
given value of magnetic field $H$. It is convenient to consider the
function $T_{D}=T_{D}(\mu_{0}H)$, defined by the
Eq.~(\ref{eq:critical curves}) at $T=0K$. Solving the
Eq.~(\ref{eq:critical curves}) at $T=0K$, one can obtain

\begin{equation} \label{eq:Dingle temperature}
T_{D}(\mu_{0}H)=\lambda(\mu_{0}H)\ln{ \frac{(2\mathcal{A}_{0}\hbar)^{2}k_{1}}{\pi^{3}m_{c}(\mu_{0}H)^{2}}Q(\mu_{0}H)} \\
\end{equation}

\begin{figure}
  \includegraphics[width=0.4\textwidth]{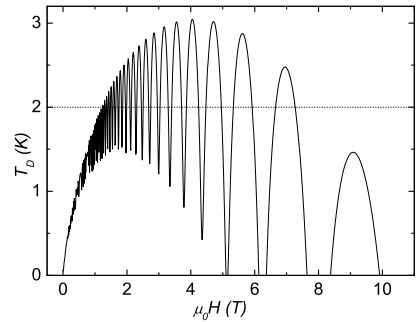}
\caption{\ Shown is the field dependence of the Dingle temperature
$T_{D}=T_{D}(\mu_{0}H)$ at $T=0$. Depending on the impurity of the
sample, there is a set of intervals in the applied magnetic field
with possible existence of the Condon domain structure. The dash
line corresponds to the value of $T_{D}=2K$, it crosses the curve
$T_{D}=T_{D}(\mu_{0}H)$ in discrete number of points which divide
the whole range of magnetic field into alternating intervals. With
increase of the field $H$ the uniform phases are replaced by
non-uniform ones with the periodicity $P(1/\mu_{0}H)=1/\Delta F$. In
calculation we put $\beta=0.15$.} \label{Dingle temperature}
\end{figure}

The results of numerical calculations of the function
$T_{D}=T_{D}(\mu_{0}H)$ (\ref{eq:Dingle temperature}) are shown in
Fig.~\ref{Dingle temperature}.

The function $T_{D}=T_{D}(\mu_{0}H)$ is characterized by oscillatory
dependence on the magnetic field $H$, which is a consequence of the
beats resulting from two close fundamental frequencies. It is
interesting to compare this dependence with analogous one for
silver. In case of silver \cite{Gordon2}, at given impurity of the
sample, e.g. fixed Dingle temperature $T_{D}$, the values of
external magnetic field with possible existence of Condon domain
structure belong to the interval $H_{<}(T_{D}<H<H_{>}(T_{D}))$. For
beryllium the whole interval of quantizing magnetic field consists
of the set of the intervals
$H^{(i)}_{<}(T_{D}<H<H^{(i)}_{>}(T_{D}))$ with alternating uniform
and non-uniform phases. The number of the intervals, where the
non-uniform phase exists, decreases with increase of Dingle
temperature $T_{D})$, collapsing at some value $T^{(max)}_{D}$,
depending on parameter $\beta$. Thus, $T^{(max)}_{D}\approx 3.1K$
for $\beta=0.15$. For $T_{D}=2K$ the width of the interval of the
existence of the non-uniform phase, defined as $\Delta
H^{i}=H^{(i)}_{>}-H^{(i)}_{<}$, increases with increasing the
external magnetic field $H$. One can show that the maximums of the
function appear periodically on the scale of inverse magnetic field
with the period inversely proportional to the discrepancy of the two
fundamental frequencies, corresponding to two extreme cross sections
of 'cigar':
\begin{equation} \label{eq:period}
P(\frac{1}{\mu_{0}H})= \frac{1}{\Delta F} \\
\end{equation}
where $\Delta F=F_{h}-F_{w}$. With $F_{h}=970.9T$ and $F_{w}=942.2T$
\cite{Egorov}, we obtain $\Delta F=28.7T$. This can be verified by
experiment on observation of non-uniform phase.

\begin{figure}
  \includegraphics[width=0.4\textwidth]{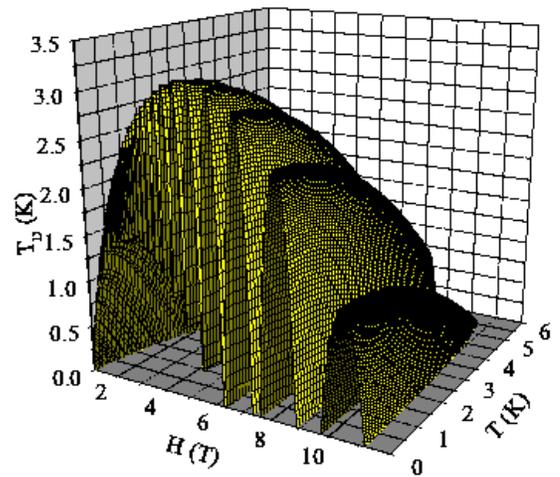}
\caption{\ Three dimensional phase diagrams in coordinates
$T-H-T_{D}$ for beryllium. Above the surface
$T_{D}=T_{D}(T,\mu_{0}H)$ the uniform phase exists, while below this
surface the non-uniform phase (Condon domain structure) is formed.}
\label{Phase diagrams}
\end{figure}

\begin{figure}[b]
  \includegraphics[width=0.4\textwidth]{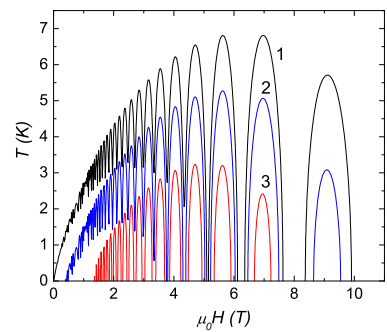}
\caption{\ Envelope of phase diagrams for beryllium $T-H$ (color
online), estimated at three different Dingle temperatures: $T_{D}=0,
1K$ and $2.1K$ (curves $1$) (black), $2$ (blue) and $3$ (red)
correspondingly. The position of the maximums is not influenced by
$T_{D}$, while the values of the maximums depend strongly on
$T_{D}$, decreasing with increase of $T_{D}$.} \label{2D}
\end{figure}

To construct the phase diagrams for beryllium we put $a=1$ in
Eq.~(\ref{eq:reduced amplitude}). The $T-H$ curves in
Fig.~\ref{Phase diagrams} form the geometric place of points which
correspond to diamagnetic phase transition at different Dingle
temperatures. Outside the shell-like surface $T_{D}=T_{D}(T,H)$ the
uniform phase takes a place, while the ordered phase is located
below this surface. At given Dingle temperature $T_{D}$ the field
dependence of the critical temperature $T=T(\mu_{0}H)$ has maximums.
The analysis of the phase diagrams estimated in the framework of
Eq.~(\ref{eq:model}) shows that the position of the maximums depends
entirely on the difference between fundamental frequencies,
corresponding to two extreme cross sections of the Fermi surface,
while the values of these maximums is defined by the impurity of the
sample, e.g. the values of Dingle temperature $T_{D}$, and also by
the parameter $\beta<0.5$, which is the characteristic of the
transition range between two extreme cross sections of Fermi sheet.
The analysis shows that the values of the maximums grows with
decrease of the Dingle temperature $T_{D}$ or increase of $\beta$.
The results of numerical calculations of the phase diagrams for
different Dingle temperatures $T_{D}=0, 1$ and $2.1K$ are
illustrated by Fig.~\ref{2D}, which shows that the positions of the
maximums of the curve $T-H$ are not affected by the Dingle
temperature, but the values of the maximums depend strongly on the
$T_{D}$. The increase of $T_{D}$ results in decrease of the values
of the maximums and their collapses. Thus, the last maximum, located
at for $T_{D}=0 K$, disappears when $T_{D}$ grows to $2.1K$.

\begin{figure}
  \includegraphics[width=0.4\textwidth]{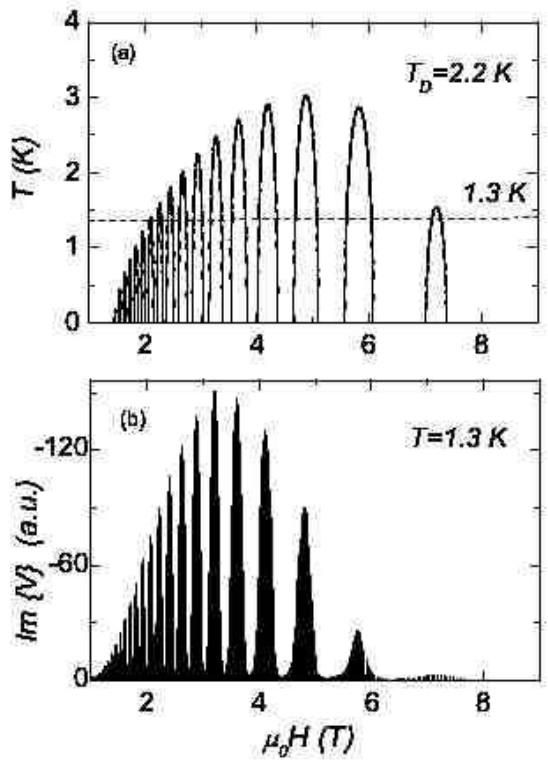}
\caption{\ Comparison computed phase diagrams with the data
\cite{Kramer2}. Typical critical curve $T=T(\mu_{0}H)$ at given
Dingle temperature $T_{D}=2.2K$ is shown ($a$). Figure ($b$)
represents the experimental data \cite{Kramer2} on investigation of
the irreversible effects in beryllium, which are the results of
formation of Condon domain structure. The output signal is shown in
arbitrary units. There is a good agreement between the intervals of
existence of the non-uniform phase, estimated in the framework of
(\ref{eq:model}), and the appearance of the observed in
\cite{Kramer2} non-linear response.} \label{Comparison}
\end{figure}

Resent experimental results on observation of irreversible effects
in beryllium \cite{Kramer2}, as well as giant parametric enhancement
of non-linear effects \cite{Tsindlekht}, allows us to verify the
phase diagrams calculated in the framework of the assumptions
(\ref{eq:model}). Undoubtedly, both of the observed effects, e.g.
irreversibility \cite{Kramer2} and strong non-linear dynamics
\cite{Tsindlekht} are closely related to the electron instability at
given temperature in increasing applied magnetic field. This
instability results in the phase transition from uniform state to
non-uniform one with formation of Condon domain structure. Both
results \cite{Kramer2}-\cite{Tsindlekht} indicate on the existence
of the large scale periodicity of the measured effects relative to
inverse magnetic field (see, Eq.~(\ref{eq:period})), which is
additional one to the usual small scale periodicity of dHvA effect.
Undoubtedly, the large scale periodicity of above-mentioned effects,
measured at given temperature ($T=1.3 K$ \cite{Kramer2} and $T=1.2
K$ \cite{Tsindlekht}) is the result of the periodicity in the
appearance of the non-uniform state and, therefore, can serve as an
instrument for verification of the theoretical phase diagrams.
Comparison between the theoretical phase diagrams calculated on the
basis of the assumption \ref{eq:model} is illustrated by
Fig.~\ref{Comparison}. In Fig.~\ref{Comparison}($a$) the calculated
diamagnetic phase transition temperature is plotted as a function of
applied magnetic field for Dingle temperature $T_{D}=2.2K$. In
calculation of phase diagram we put $\beta=1.15$. In
Fig.~\ref{Comparison}($b$) the amplitude of the third harmonic
\cite{Kramer2} (in arbitrary units), arising in the conditions of
electron instability, as a function of magnetic field is depicted.
It is evident from the comparison the theoretical phase diagram and
data \cite{Kramer2}, measured at $T=1.3K$, that non-linear response
appears at the intervals of applied magnetic field, corresponding to
the formation of the Condon domain structure. The inverse large
scale period $\Delta F=28.7T$ \ref{eq:period} gives a good agreement
between theory and experiment. Unfortunately, the relevant to our
calculation measurements in \cite{Kramer2} were made only at fix
temperature $T=1.3K$. The measurements of the temperature dependence
of the non-linear response (together with the field dependencies)
would make it possible to reconstruct the complete phase diagrams
$T=T(\mu_{0}H)$ for beryllium.

\section{Conclusions}

We developed the model of slightly corrugated Fermi surface of
beryllium with taking into account the negligible small curvature of
the extreme cross sections of the Fermi surface sheet relevant to
observation of dHvA oscillations. The calculated in the framework of
our theory phase diagrams for beryllium are in a good agreement with
available experimental data on observation of Condon domain
instability.

The estimated phase diagrams reveal large scale periodicity on
reciprocal magnetic field. The inverse period does not depends on
the impurity of the sample and is related to the discrepancy of the
two fundamental frequencies $\Delta F=28.7T$ Eq.~(\ref{eq:period}) ,
corresponding to two extreme cross section of 'cigar' like Fermi
surface of beryllium, e.g. 'waist' and 'hips'. The magnetic field
dependencies of recent experimental data on observation of
non-linear effects in beryllium \cite{Kramer2}-\cite{Tsindlekht}
have a simple explanation on the basis of the estimated diamagnetic
phase diagrams. We hope that the theoretical results will stimulate
further experiments on investigation of electron instability in
strongly correlated electron systems. In particular, the
measurements of the temperature dependencies of the non-linear
effects in the samples of different impurity will allow constructing
the complete phase diagrams.

\begin{acknowledgments}
We are grateful to R. Kramer and I. Sheikin for many valuable
discussions. We express our deep gratitude to P. Wyder for his
interest in this work.
\end{acknowledgments}

\end{document}